\documentclass[preprint,showpacs,titlepage,aps,prd,
tightenlines,
amsmath,byrevtex,nofootinbib]{revtex4}

\usepackage{graphicx}

\newcommand{\dMsq}{\Delta m^2_{ \text{ren} }}

\def\lsim{\raise0.3ex\hbox{$\;<$\kern-0.75em\raise-1.1ex
\hbox{$\sim\;$}}}
\def\gsim{\raise0.3ex\hbox{$\;>$\kern-0.75em\raise-1.1ex
\hbox{$\sim\;$}}}

\begin{document}

\title{Flavor-Universal Form of Neutrino Oscillation Probabilities \\ in Matter}\thanks{\it Written version of the talk presented at NuFact15, 10-15 Aug 2015, Rio de Janeiro, Brazil [C15-08-10.2] under the title ``Compact Formulas for Neutrino Oscillation Probabilities in Matter''.}

\vskip 16mm 

\author{Hisakazu~Minakata$^{a, b}$}
\email{hisakazu.minakata@gmail.com}
\affiliation{
$^a$Instituto de F\'{\i}sica, Universidade de S\~ao Paulo, C.\ P.\
66.318, 05315-970 S\~ao Paulo, Brazil \\
$^b$Yachay Tech University, School of Physical Sciences and \\ Nanotechnology, %Hacienda, San Jos\'e s/n y Proyecto Yachay, 
San Miguel de Urcuqu\'i, Ecuador  }
\thanks{Address after April 2016.}

%\date{\today}
\date{December 20, 2015}

\vglue 1.6cm

\begin{abstract}
We construct a new perturbative framework to describe neutrino oscillation in matter with the unique expansion parameter $\epsilon$, which is defined as $\Delta m^2_{21} / \dMsq$ with the renormalized atmospheric $\dMsq \equiv \Delta m^2_{31} - s^2_{12} \Delta m^2_{21}$. It allows us to derive the maximally compact expressions of the oscillation probabilities in matter to order $\epsilon$ in the form akin to those in vacuum. This feature allows immediate physical interpretation of the formulas, and facilitates understanding of physics of neutrino oscillations in matter. Moreover, quite recently, we have shown that our three-flavor oscillation probabilities $P(\nu_\alpha \rightarrow \nu_\beta)$ in all channels can be expressed in the form of universal functions of $L/E$. The $\nu_e$ disappearance oscillation probability $P(\nu_e \rightarrow \nu_e)$ has a special property that it can be written as the two-flavor form which depends on the single frequency. This talk is based on the collaborating work with Stephen Parke \cite{Minakata:2015gra}.

\end{abstract}
\pacs{14.60.Lm, 14.60.Pq }

\maketitle

\section{Introduction}
\label{sec:introduction}

Do we understand neutrino oscillation? Most experimentalists and most theorists would agree to answer {\em ``Yes we do''}. There is a simple way to derive, in vacuum and in matter, the oscillation probability and apparently it describes well the available experimental data. 

However, I want to point out that not every aspect of theory of neutrino oscillation has been tested experimentally. For example, to my knowledge,  

\begin{itemize}
\item
No one observed neutrinos directly in their mass eigenstates as a whole.\footnote{
%%%%%%%%%%%%% footnote %%%%%%%%%%%%%
One may argue that observation of $^8$B solar neutrinos detect $\nu_{2}$ in a good approximation. But, it still detects $\nu_{e}$ component of $\nu_{2}$ if one uses CC reaction. Detection by NC reaction does not alter this situation, because a particular component of $\nu_{2}$ causes the reaction in each time.  
}
It probably requires detection of neutrinos by gravitational effects, and in this context, cosmological observation is likely to be the first runner to achieve the goal, see e.g., \cite{Ade:2015xua}. 

\item
Nobody observed the effect of neutrino's wave packet. See for example \cite{Akhmedov:2012uu} for a recent treatment. If someone could develop technology which has sensitivity to the size or shape of the wave packet, then it would become possible to see it. If the time resolution of detector is improved dramatically, in principle, it may allow us to detect the effect of superluminal neutrinos due to oscillation-driven modification of shape of the wave packet in flight \cite{Minakata:2012kg}. 

\end{itemize}

I said in the above that ``there is a simple way to derive the oscillation probability in vacuum and in matter''. In fact, this comment is only true for the regime in which single-$\Delta m^2$ dominance approximation applies, and the things are quite different beyond it. Now, the various neutrino experiments entered into the regime where the three-flavor effects become important. Or, precision of measurement became so high that it has sensitivity to the sub-leading effects. See e.g., \cite{Gando:2013nba,Okumura:2015dna}. The accelerator neutrino experiment \cite{Abe:2015oar,Bian:2015opa} is the best example for the former because the CP phase effect, not only $\sin \delta$ but also $\cos \delta$ effect, is the genuine three-flavor effect. This is best understood by the general theorems derived in Refs.~\cite{Naumov:1991ju,Harrison:1999df} ($\sin \delta$ terms) and \cite{Asano:2011nj} ($\cos \delta$ terms). 

Let us focus on the accelerator neutrino experiment because it will play a major role in observing the CP phase effect in a robust way \cite{Abe:2015zbg,Akiri:2011dv}. In the regime where the three-flavor effect is important our theoretical understanding of the neutrino oscillation probability is not quite completed in my opinion. Let me first try to convince the readers on this point. For pedagogical purpose, I start from neutrino oscillation in vacuum. If you want to know the key point go directory to section~\ref{sec:matter-simple}.

\section{The oscillation probability in vacuum is simple}
\label{sec:vacuum}

The neutrino oscillation probability in vacuum is simple. If only two generations of neutrinos ($\nu_e$ and $\nu_\mu$) exist it takes the form 
\begin{eqnarray}
P(\nu_\mu \rightarrow \nu_e)  &=&
\sin^2 2 \theta ~\sin^2 \left( \frac{ \Delta m^2 L }{4E} \right) 
\label{eq:P-ee-vacuum}
\end{eqnarray}
where $\theta$ denotes the mixing angle and $\Delta m^2 = m^2_{2} - m^2_{1}$. The variable $\frac{ \Delta m^2 L }{2E}$ in the sine function is nothing but the phase difference between the mass eigenstates $\nu_2$ and $\nu_1$ which is developed when neutrinos travelled a distance $L$. Whereas the strength of the oscillation is determined by the transition amplitude $\sin 2 \theta$.

In nature the three-generation neutrinos exist, $\nu_\alpha$ ($\alpha = e, \mu, \tau$) in the favor basis and $\nu_{i}$ ($i=1, 2, 3$) in the mass eigenstate basis. Let us define the MNS lepton flavor mixing matrix \cite{Maki:1962mu} as $\nu_\alpha= U_{\alpha i} \nu_{i}$. Then, the neutrino oscillation probability has richer structure with more terms with different characteristics: 
\begin{eqnarray}
P(\nu_\beta \rightarrow \nu_\alpha) &=& 
\delta_{\alpha \beta}  - 4 \sum_{j > i}  {\mbox Re} [U_{\alpha 1} U_{\beta 1}^* U_{\alpha 2}^* U^{ }_{\beta 2}]  
\sin^2 \frac{ \Delta m^2_{ji} L}{ 4E } 
\nonumber\\
&+&  %\hspace{-1cm}+~
8 ~{\mbox Im} [U_{\alpha 1} U_{\beta 1}^* U_{\alpha 2}^* U^{ }_{\beta 2}]  
~\sin \frac{ \Delta m^2_{32} L}{ 4E } 
\sin \frac{ \Delta m^2_{21} L}{ 4E } 
\sin \frac{ \Delta m^2_{31} L}{ 4E }.  
\label{eq:P-ba-general-vacuum}
\end{eqnarray}
In addition to the proliferation of the conventional term that appear in (\ref{eq:P-ee-vacuum}) due to the three mass-squared differences, there arises a universal CP and T violating term, the last one in (\ref{eq:P-ba-general-vacuum}). The term is suppressed by the two small factors, the Jarlskog factor \cite{Jarlskog:1985ht} ${\mbox Im} [U_{\alpha 1} U_{\beta 1}^* U_{\alpha 2}^* U^{ }_{\beta 2}] = c_{12} s_{12} c_{23} s_{23} c^2_{13} s_{13} \sin \delta < 0.035$, and $\frac{ \Delta m^2_{21} L}{ 4E } \sim \frac{ \Delta m^2_{21} }{ \Delta m^2_{31} } \simeq 0.03$ assuming that $\frac{ \Delta m^2_{31} L}{ 4E } \sim 1$. They both indicates that the CP violation is a genuine three flavor effect.   

\section{The oscillation probability in matter is complicated}
\label{sec:matter-complicated}

It is well known that under the constant matter density approximation the neutrino oscillation probability in matter can be expressed in the form in (\ref{eq:P-ba-general-vacuum}), but with replacement 
\begin{eqnarray}
\Delta m^2_{ji} \rightarrow \lambda_{j} - \lambda_{i}, 
\hspace{10mm}
U_{\alpha i} \rightarrow V_{\alpha i}, 
\label{eq:replacement}
\end{eqnarray}
where $V_{\alpha i}$ is the mixing matrix in matter defined as $\nu_\alpha= V_{\alpha i} \nu_{i}^{\text{m}}$ with $\nu_{i}^{\text{m}}$ ($i=1,2,3$) being the mass eigenstate in matter. $\frac{\lambda_{i}}{2E}$ denote the eigenvalues of the Hamiltonian in matter, 
\begin{eqnarray}
H= 
\frac{ 1 }{ 2E } \left\{ 
U \left[
\begin{array}{ccc}
0 & 0 & 0 \\
0 & \Delta m^2_{21}& 0 \\
0 & 0 & \Delta m^2_{31} 
\end{array}
\right] U^{\dagger}
+
\left[
\begin{array}{ccc}
a & 0 & 0 \\
0 & 0 & 0 \\
0 & 0 & 0
\end{array}
\right] 
\right\}, 
\label{eq:hamiltonian}
\end{eqnarray}
where $a \equiv 2\sqrt{2} G_F N_e E$ is the Wolfenstein matter potential \cite{Wolfenstein:1977ue} with electron number density $N_e$ and the Fermi constant $G_F$. The Hamiltonian governs the evolution of neutrino states as $i \frac{d}{dx} \nu = H \nu$.

Then, you may say that the oscillation probability in matter, Eq.~(\ref{eq:P-ba-general-vacuum}) with the replacement (\ref{eq:replacement}), is structurally very simple. It is true. Even more amazingly one can obtain the exact expressions of the $V$ matrix elements \cite{Zaglauer:1988gz,Kimura:2002wd}. However, you will be convinced if you look into the resulting expressions by yourself that they are terribly complicated, and it is practically impossible to read off some physics from the expressions. Sorry, I have no space here to introduce you the beautiful method for calculating the $V$ matrix elements introduced in Ref.~\cite{Kimura:2002wd}, and demonstrate the complexity of the resultant expression.

\subsection{We need perturbation theory, but it is {\em not} enough}

Here is a natural question you may raise: 
``Isn't it possible to compute the eigenvalues $\lambda_{i}$ and $V_{\alpha i}$ perturbatively?\footnote{
%%%%%%%%%%%%%%% footnote %%%%%%%%%%%%%
In fact, it is a highly nontrivial question why the expansion of the exact expression of $\lambda_{i}$ and $V$ matrix elements by the small parameter does not work. This question is briefly addressed in \cite{Minakata:2015gra}.
}
If you take this way you must be able to obtain much simpler analytic expressions of the oscillation probabilities.'' Yes, of course you can. But, when you engage this business you discover that the eigenvalues $\lambda_{i}$ receives the first order corrections. When you expand by the small parameters your formulas for the oscillation probabilities do not remain to the structure-revealing form (\ref{eq:P-ba-general-vacuum}). Usually you obtain proliferation of terms, and the situation becomes much worse when you go to higher orders.
This is the characteristic feature of the expressions obtained by the perturbative frameworks so far examined, to our understanding.\footnote{
%%%%%%%%%%%%% footnote %%%%%%%%%%%%%%%
I hope you understand that this comment is not to hurt the previous authors' efforts devoted to understand the neutrino oscillations by developing the various perturbative schemes. In talking about the proliferation of terms, in fact, the present author was very good at producing lengthy formulas: He is proud of deriving the longest formula for $P(\nu_e \rightarrow \nu_\mu)$ expanded to third order in $\varepsilon \equiv \frac{\Delta m^2_{21}}{\Delta m^2_{31}}$, $\sin \theta_{13}$, and even including the NSI parameters to the same order, which spanned 3 pages when it is explicitly written. See arXiv version 1 of \cite{Kikuchi:2008vq}. If you are interested in seeing the other (but much less pronounced) examples see section 3.3.6 in Ref.~\cite{Minakata:2015gra}.
}

Since it is very hard to collect all the relevant references in which the various perturbative frameworks are developed, please look at the bibliography in \cite{Minakata:2015gra,Asano:2011nj,Kikuchi:2008vq} for an incomplete list of references, from which you can start your own search. 

Then, the immediate question would be 
``Can't you construct perturbation theory in which the first order corrections to the eigenvalues $\lambda_{i}$ are absent?''. If we can, the proliferation of terms is avoided and the simple structure of the oscillation probabilities in (\ref{eq:P-ba-general-vacuum}) is maintained to first order in the expansion parameter. The answer to the above question is {\em Yes} and this is what we did in Ref.~\cite{Minakata:2015gra}. 

\section{The oscillation probability in matter can be made extremely simple and compact}
\label{sec:matter-simple}

The next question we must ask is then: How can we make the first order correction to the eigenvalues $\lambda_{i}$ vanishes? There is a simple way to make it happen. That is, if we choose the decomposition of the Hamiltonian into the unperturbed and the perturbed parts correctly, then it is automatic. For concreteness I want to describe how it happens in the perturbative framework we have developed in \cite{Minakata:2015gra}.

We first go to the tilde basis $\tilde{H} = U_{23}^{\dagger} H U_{23}$. Then, we decompose $\tilde{H}$ as 
$\tilde{H} (x) = \tilde{H}_{0} (x) + \tilde{H}_{1} (x)$:
\begin{eqnarray} 
\tilde{H}_{0} (x) &=& 
\frac{ \dMsq }{ 2E } \left\{
\left[
\begin{array}{ccc}
\frac{ a }{ \dMsq}  + s^2_{13} & 0 & c_{13} s_{13} \\
0 & 0 & 0 \\
c_{13} s_{13}  & 0 & c^2_{13} 
\end{array}
\right] 
+
\epsilon 
\left[
\begin{array}{ccc}
s^2_{12} & 0 & 0 \\
0 & c^2_{12} & 0 \\
0 & 0 & s^2_{12} 
\end{array}
\right]  \right\}, 
\label{eq:H-tilde-zeroth}
\\[3mm]
\tilde{H}_{1} (x) &=& 
\epsilon c_{12} s_{12} \frac{ \dMsq }{ 2 E }
\left[
\begin{array}{ccc}
0 & c_{13} & 0 \\
c_{13} & 0 & - s_{13}  \\
0 & - s_{13} & 0 
\end{array}
\right], 
\label{eq:H-tilde-first}
\end{eqnarray} 
where 
\begin{eqnarray} 
\dMsq \equiv \Delta m^2_{31} - s^2_{12} \Delta m^2_{21}, 
\hspace{10mm} 
\text{and}
\hspace{10mm} 
\epsilon \equiv \Delta m^2_{21}/\dMsq. 
\label{eq:def-Dm2ren-epsilon}
\end{eqnarray} 
The vanishing diagonal terms in the perturbed Hamiltonian (\ref{eq:H-tilde-first}) guarantees the absence of the first-order corrections to the eigenvalues. Then, we can obtain the structure-revealing form of the oscillation probabilities in matter, Eq.~(\ref{eq:P-ba-general-vacuum}) with the replacement (\ref{eq:replacement}), to first order in $\epsilon$. Notice that use of the renormalized $\Delta m^2_{atm}$ defined in (\ref{eq:def-Dm2ren-epsilon}) makes the form of the tilde-Hamiltonian very neat. Because of the use of the unique expansion parameter $\epsilon$ provided by nature, we have named our perturbative framework as ``renormalized helio-perturbation theory'' \cite{Minakata:2015gra}.

\section{Universal form of neutrino oscillation probabilities in matter}
\label{sec:universal}

This is not the end of the story. We have observed the following two  ``unexpected'' new features. If we write down the disappearance oscillation probability $P(\nu_e \rightarrow \nu_e)$ in our renormalized helio-perturbation theory, it is extremely simple. To order $\epsilon$ it reads
\begin{eqnarray}
P(\nu_e \rightarrow \nu_e)  &=&
1 -  \sin^2 2\phi  ~\sin^2 \frac{ (\lambda_{+} - \lambda_{-} ) L }{4E} 
\label{eq:P-ee-matter}
\end{eqnarray}
where $\lambda_{-}, \lambda_{0}, \lambda_{+} $ denote the three eigenvalues of $2E \tilde{H}_{0}$. $\phi$, the mixing $\theta_{13}$ in matter, is given by 
\begin{eqnarray} 
\cos 2 \phi &=& 
\frac{ \dMsq \cos 2\theta_{13} - a }{ \lambda_{+} - \lambda_{-} },
\nonumber \\
\sin 2 \phi &=& \frac{ \dMsq \sin 2\theta_{13} }{ \lambda_{+} - \lambda_{-} }.
\label{eq:cos-sin-2phi}
\end{eqnarray}
Compare the expression in (\ref{eq:P-ee-matter}) to the vacuum formula in (\ref{eq:P-ee-vacuum}). So similar! Notice that, though extremely compact, it contains all-order contributions of both $s_{13}$ and $a$. 

The leading order $\epsilon^0$ term in the appearance channel probability $P(\nu_e \rightarrow \nu_\mu)$ calculated to order $\epsilon$ is also governed by the particular frequency $\lambda_{+} - \lambda_{-}$: 
\begin{eqnarray}
&& P(\nu_e \rightarrow \nu_\mu) 
\nonumber \\
&=& 
\left[ 
s^2_{23} \sin^2 2 \theta_{13}
+ 
4 \epsilon   
J_r \cos \delta 
\left\{ \frac{ (\lambda_{+} - \lambda_{-}) - ( \dMsq - a ) }{  ( \lambda_{+} - \lambda_{0} ) } \right\}
\right]
 \left(\frac{\dMsq}{ \lambda_{+} - \lambda_{-} }\right)^2 \sin^2 \frac{ (\lambda_{+} - \lambda_{-}) L}{ 4E } 
\nonumber \\[2mm]
&+& 8 \epsilon  
J_r 
\frac{ (\dMsq)^3 }{ ( \lambda_{+} - \lambda_{-} ) ( \lambda_{+} - \lambda_{0} ) ( \lambda_{-} - \lambda_{0} ) }
\sin \frac{ (\lambda_{+} - \lambda_{-}) L}{ 4E } 
\sin \frac{ (\lambda_{-} - \lambda_{0}) L}{ 4E} 
\cos \left( \delta - \frac{ (\lambda_{+} - \lambda_{0}) L}{ 4E } \right)
\nonumber \\
\label{eq:P-emu-matter}
\end{eqnarray}
where $J_r$, the reduced Jarlskog factor, is defined as 
\begin{eqnarray} 
J_r \equiv c_{12} s_{12} c_{23} s_{23} c^2_{13} s_{13}.
\label{eq:Jarlskog-def}
\end{eqnarray}
This expression (\ref{eq:P-emu-matter}) is quite compact, despite that it contains all-order contributions of $s_{13}$ and $a$. In particular, it keeps the similar structure as the one derived by the Cervera {\it et al.} \cite{Cervera:2000kp}, which retains terms of order $\epsilon^2$ but is expanded by $s_{13}$ only up to second order. 

Furthermore, quite recently, we have observed that the first-order formulas for the oscillation probabilities have the flavor-universal (up to $\theta_{23}$-dependent coefficient) expressions. Namely, $P(\nu_\alpha \rightarrow \nu_\beta)$ (including the $\nu_{e}$ sector) can be written in a universal form:
%
%{\small
\begin{eqnarray}
&& P(\nu_\alpha \rightarrow \nu_\beta) =  \delta_{\alpha \beta}
\nonumber \\ 
&& \hspace*{-0.5cm}
+ ~4\left[ 
 \{A^{\alpha \beta}_{+-}\}~s^2_\phi c^2_\phi
+ \epsilon ~\{B^{\alpha \beta}_{+-}\}   \left(J_r \cos \delta \right) 
\frac{ (\dMsq)^2  
\left\{ ( \lambda_{+} - \lambda_{-} ) - ( \dMsq - a ) 
\right\} }{ ( \lambda_{+} - \lambda_{-} )^2 ( \lambda_{+} - \lambda_{0} ) } 
\right]
~\sin^2 \frac{ (\lambda_{+} - \lambda_{-}) L}{ 4E } 
\nonumber \\[2mm] 
&& \hspace*{-0.5cm}
+ 
~4\left[
\{A^{\alpha \beta}_{+0}\}~c_\phi^2 
~+  \epsilon ~\{B^{\alpha \beta}_{+0}\} \left(J_r \cos \delta/c^2_{13} \right) 
\frac{ \dMsq \left\{ ( \lambda_{+} - \lambda_{-} ) - ( \dMsq + a ) \right\} }{ ( \lambda_{+} - \lambda_{-} ) ( \lambda_{+} - \lambda_{0} ) } 
\right]
~\sin^2 \frac{ (\lambda_{+} - \lambda_{0}) L}{ 4E} 
\nonumber \\[2mm] 
&& \hspace*{-0.5cm}
+ 
~4 \left[
\{A^{\alpha \beta}_{-0}\}~ s_\phi^2 
~+\epsilon ~\{B^{\alpha \beta}_{-0}\} \left(J_r \cos \delta/c^2_{13} \right)
\frac{ \dMsq \left\{ ( \lambda_{+} - \lambda_{-} ) + ( \dMsq + a ) \right\} }{ ( \lambda_{+} - \lambda_{-} ) ( \lambda_{-} - \lambda_{0} ) } 
\right]
~\sin^2 \frac{ (\lambda_{-} - \lambda_{0}) L}{ 4E } 
\nonumber\\[2mm]
&& \hspace*{-0.5cm}
 + ~8 \epsilon ~J_r  
~\frac{ (\dMsq)^3 }{ ( \lambda_{+} - \lambda_{-} ) ( \lambda_{+} - \lambda_{0} ) ( \lambda_{-} - \lambda_{0} ) }
~\sin \frac{ (\lambda_{+} - \lambda_{-}) L}{ 4E } 
~\sin \frac{ (\lambda_{-} - \lambda_{0}) L}{ 4E } 
\nonumber \\[2mm]
& & \hspace*{3cm}  \times
\left[
~\{C^{\alpha \beta}\}  ~\cos \delta ~\cos \frac{ (\lambda_{+} - \lambda_{0}) L}{ 4E} 
+ ~\{S^{\alpha \beta}\} ~ \sin \delta ~\sin \frac{ (\lambda_{+} - \lambda_{0}) L}{ 4E } 
~\right]. 
\label{eq:P-albe-sec3}
\end{eqnarray}
%}
%
\begin{table}[t]
%\hspace*{-1.5cm} 
\begin{tabular}{|c|c|c|c|c|c|c|}
 \hline    & & & & & & \\
 &  $\nu_e \rightarrow  \nu_e$  &     $\begin{array}{c} \nu_e \rightarrow \nu_\mu \\  \nu_\mu \rightarrow \nu_e \end{array}  $  &  
 $\begin{array}{c} \nu_e \rightarrow \nu_\tau \\  \nu_\tau \rightarrow \nu_e \end{array}  $   &  $\begin{array}{c} \nu_\mu \rightarrow \nu_\tau \\  \nu_\tau \rightarrow \nu_\mu \end{array}  $   &  $ \nu_\mu \rightarrow \nu_\mu$  &  $\nu_\tau \rightarrow \nu_\tau $   \\[6mm] \hline
  Order $\epsilon^0$:  & & & & & & \\[1mm]
  $A^{\alpha \beta}_{+-}$ & -1 & $\sin^2 \theta_{23} $ &  $\cos^2 \theta_{23} $ &  $-\sin^2 \theta_{23} \cos^2 \theta_{23} $ &
  $-\sin^4 \theta_{23} $ & $-\cos^4 \theta_{23} $ \\[1mm]
  $A^{\alpha \beta}_{+0} =A^{\alpha \beta}_{-0}$ & 0 & 0 & 0 & $\sin^2 \theta_{23} \cos^2 \theta_{23} $  &  $-\sin^2 \theta_{23} \cos^2 \theta_{23}$ &  $-\sin^2 \theta_{23} \cos^2 \theta_{23}$ %\\[1mm]
%   $A^{\alpha \beta}_{-0}$  & 0 & 0 & 0 & $\sin^2 2 \theta_{23} $  &  $-\sin^2 2 \theta_{23} $  &$-\sin^2 2 \theta_{23} $  
\\[3mm] \hline  Order $\epsilon \cos \delta$: & & & & & & \\[1mm]
   $B^{\alpha \beta}_{+-} = C^{\alpha \beta}$ &  0 & 1 & -1 & $-\cos 2 \theta_{23}$ &  $\cos 2 \theta_{23}-1$ %$-2\sin^2 \theta_{23}$ 
   & $\cos 2 \theta_{23}+1$ \\[1mm] %$2\cos^2 \theta_{23}$ \\
   $B^{\alpha \beta}_{+0}=B^{\alpha \beta}_{-0}$  & 0 & 0 & 0 &   $-\cos 2 \theta_{23}$ &  $\cos 2 \theta_{23}$ &  $\cos 2 \theta_{23}$ 
   %\\[1mm]
%   $B^{\alpha \beta}_{-0}$  & 0 & 0 & 0 & $-\cos 2 \theta_{23}$ & $\cos 2 \theta_{23}$ & $\cos 2 \theta_{23}$ 
\\[3mm]
\hline    Order $\epsilon \sin \delta$: & & & & & & \\[1mm]
%    $C^{\alpha \beta}$ &  0 & 1 & -1 & $-\cos 2 \theta_{23}$ & $\cos 2 \theta_{23} -1$ & $\cos 2 \theta_{23} +1$  \\[1mm]
  $  S^{\alpha \beta} $ &  0  & $\pm$1 & $\mp$1 & $\pm$1 & 0 & 0 \\[3mm] \hline
  \end{tabular} 
  \caption{The values for the 5 coefficients for all oscillation channels, $\nu_\alpha \rightarrow \nu_\beta$ and  $\bar{\nu}_\alpha \rightarrow \bar{\nu}_\beta$ to be used in conjunction with eq.~(\ref{eq:P-albe-sec3}). Note that they are $0, ~\pm1$ or simple functions of $\theta_{23}$.}
\label{tab:coeff}
\end{table}
The eight coefficients $A^{\alpha \beta}_{ij}$,  $B^{\alpha \beta}_{ij}$,  $C^{\alpha \beta}$ and $S^{\alpha \beta}$ are given in Table~\ref{tab:coeff}. Notice that they are $0, ~\pm1$, or the simple functions of $\theta_{23}$. The antineutrino oscillation probabilities $P(\bar{\nu}_\alpha \rightarrow \bar{\nu}_\beta)$ can be easily obtained from the neutrino oscillation probabilities as $P(\bar{\nu}_\alpha \rightarrow \bar{\nu}_\beta: E) = P(\nu_\alpha \rightarrow \nu_\beta: - E)$. See Ref.~\cite{Minakata:2015gra} for explanation. 

We observe in Table~\ref{tab:coeff} the existence of three equalities between the coefficients
\begin{eqnarray}
A^{\alpha \beta}_{-0}= A^{\alpha \beta}_{+0},  \quad B^{\alpha \beta}_{-0}= B^{\alpha \beta}_{+0}  \quad  {\rm and} \quad  B^{\alpha \beta}_{+-}= C^{\alpha \beta} 
\label{eq:ABCids}
\end{eqnarray}
which hold due to the invariance of the oscillation probabilities under the following transformation
\begin{equation}
\phi \rightarrow \pi/2 + \phi  \quad  {\rm and} \quad  \lambda_+ \leftrightarrow \lambda_-. 
\label{eq:invar}
\end{equation}
The invariance (\ref{eq:invar}) must hold because the two cases in (\ref{eq:invar}) are both equally valid two ways of diagonalizing the zeroth-order Hamiltonian. Look at (\ref{eq:cos-sin-2phi}) to observe that the defining equations of $\phi$ are invariant under (\ref{eq:invar}). Then, the former two identities in (\ref{eq:ABCids}) trivially follow, but the last one requires use of the kinematic relationship  
\begin{eqnarray}
\sin \Delta_{+-}  \sin \Delta_{+0} \cos \Delta_{-0}  &=& \sin \Delta_{+-}  \sin \Delta_{-0} \cos \Delta_{+0}  +  \sin^2 \Delta_{+-}
\nonumber
\end{eqnarray}
where $\Delta_{ji} \equiv \frac{ (\lambda_{j} - \lambda_{i} ) L }{4E}$. 
Notice that the relation $B^{\alpha \beta}_{+-}= C^{\alpha \beta}$ needs to be satisfied only to order $\epsilon^0$ because these terms are already suppressed by $\epsilon$.

The two new features of the oscillation probabilities, the flavor-universal expressions of the oscillation probabilities $P(\nu_\alpha \rightarrow \nu_\beta)$ in (\ref{eq:P-albe-sec3}), and the extremely compact disappearance oscillation probability $P(\nu_e \rightarrow \nu_e)$ in (\ref{eq:P-ee-matter}) 
is the most remarkable outcome of our renormalized helio-perturbation theory examined to order $\epsilon$. 

\section{How accurate are our formulas?}

\begin{figure}%[htbp]
\begin{center}
\vspace{-6mm}
\includegraphics[width=0.48\textwidth]{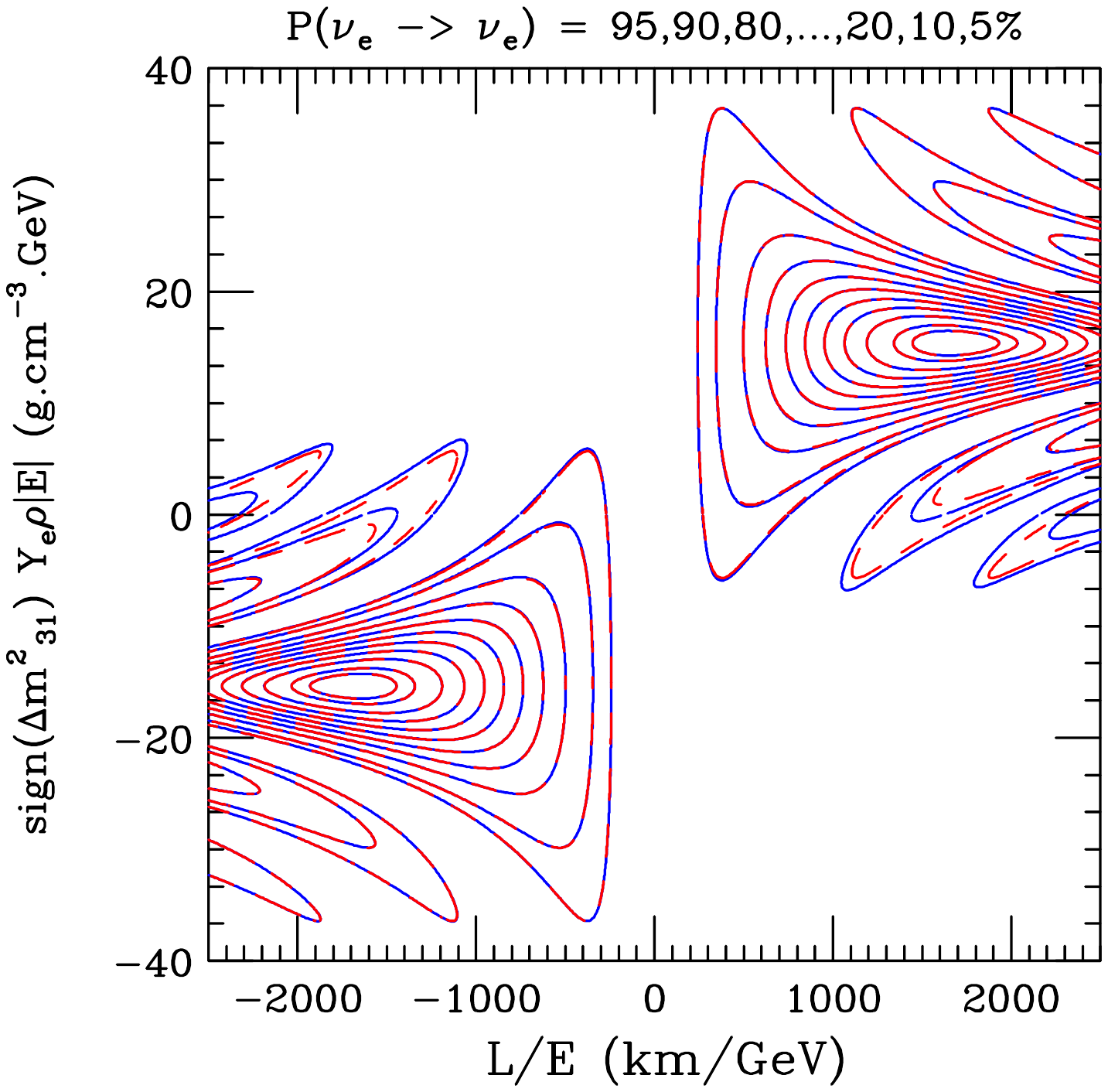}
\includegraphics[width=0.48\textwidth]{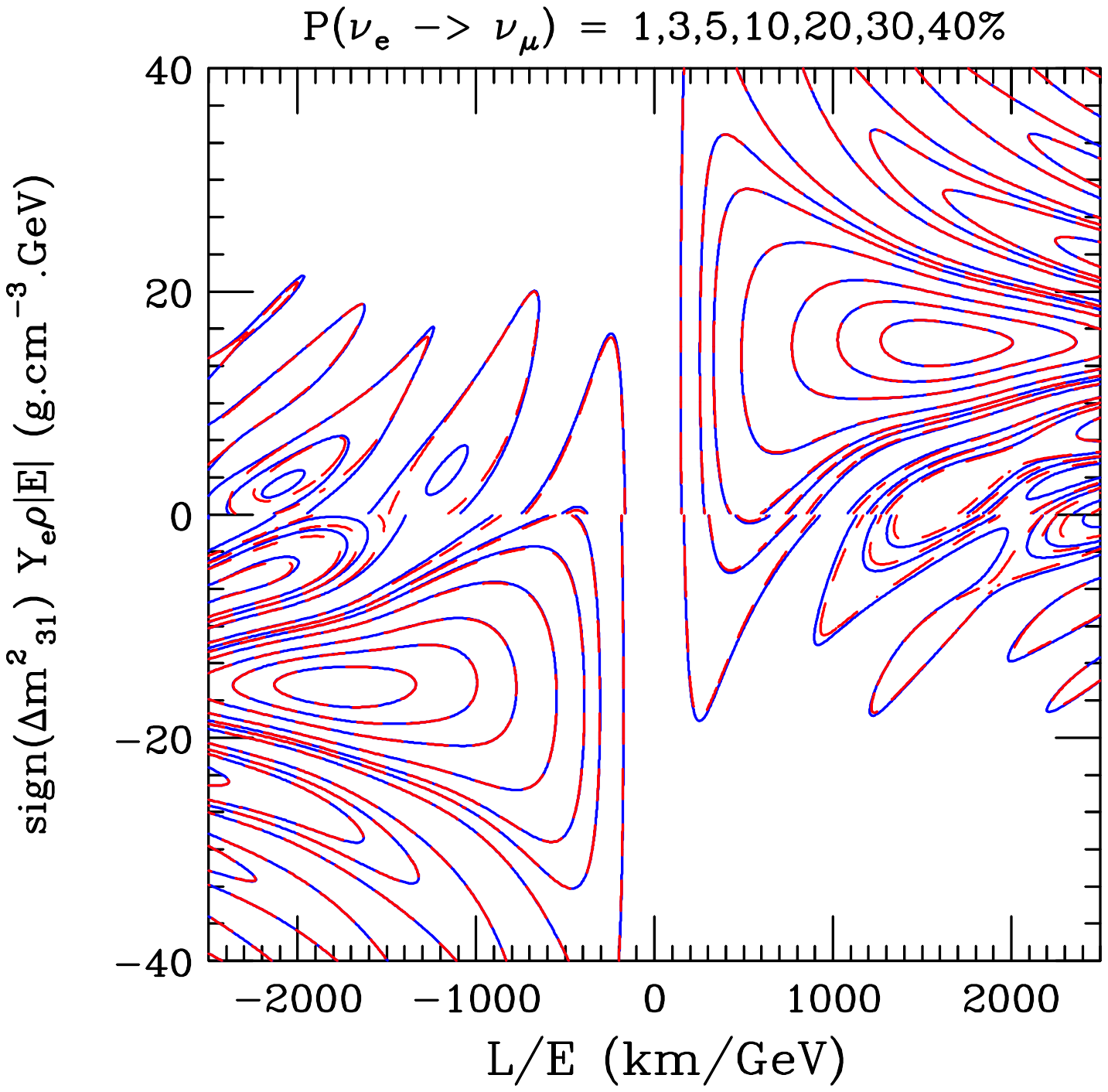}
\includegraphics[width=0.48\textwidth]{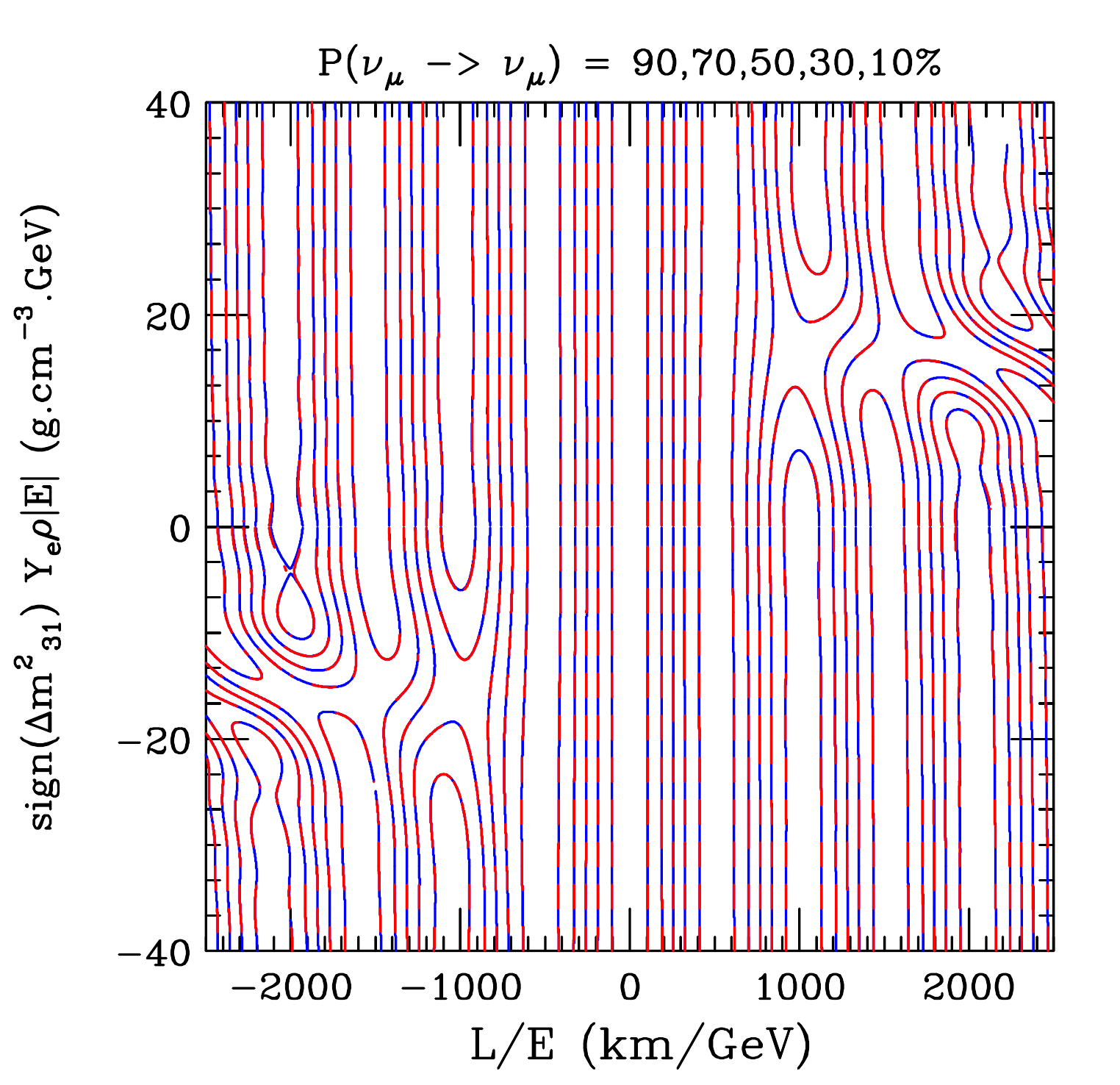}
\end{center}
\vspace{-6mm}
\caption{The iso-probability contours for the exact (solid blue) and approximate (dashed red) oscillation probabilities  for upper left, $\nu_e \rightarrow \nu_e$, upper right, $\nu_e \rightarrow \nu_\mu$ and lower, $\nu_\mu \rightarrow \nu_\mu$.
 The upper (lower) half plane is for normal ordering (inverted ordering), whereas positive (negative) L/E is for neutrinos (antineutrinos). For treatment of antineutrinos, see section~\ref{sec:universal}.
The order of the contours given in the title is determined from the line L/E=0.
The discontinued as one crosses $Y_e \rho |E| =0$ is because we are switching mass orderings at this point.
In most of parameter space the approximate and exact contours sit on top of one another so the lines appear to alternate blue-red dashed. Note that, for L/E  $>$1000 km/GeV and $|Y_e \rho E| < 5$ g cm$^{-3}$ GeV, the difference between the exact and approximate contours becomes noticeable at least for $\nu_e \rightarrow \nu_e$ and $\nu_e \rightarrow \nu_\mu$.
}
\label{fig:contours}
\end{figure}

After hearing so much advertisement such as ``structure-revealing'' or ``extremely compact'', you probably want to ask the question ``how accurate are the formulas for the oscillation probabilities?''. It is certainly a legitimate question. In Fig.~\ref{fig:contours} we present the contours of equal probability for the exact (solid blue) and the approximate (dashed red) solutions for the channels  $\nu_e \rightarrow \nu_\mu$,  $\nu_e \rightarrow \nu_e$ and $\nu_\mu \rightarrow \nu_\mu$. The right (left) half plane of each panel of Fig.~\ref{fig:contours} corresponds to the neutrino (anti-neutrino) channel. 

Overall, there is a good agreement. 
For large values of the matter potential,  $|a| > \frac{1}{3}  |\dMsq| $ we have no restrictions on  L/E to have a good approximation to the exact numerical solutions. Whereas for small  values of the matter potential,  $|a| < \frac{1}{3}  |\dMsq|$, we still need the restriction $L/E \lsim 1000$ km/GeV. The agreement between the exact and approximate formulas is worst at around the solar resonance, which is actually close to the vacuum case. The reasons for this behavior and how to interpret the drawback are discussed in \cite{Minakata:2015gra}. In the $\nu_\mu \rightarrow \nu_\mu$ channel the agreement is almost perfect due to the presence of order unity term in the oscillation probability. 

%%%%%%%%%%%%%%%% FIG 1 %%%%%%%%%%%%%%%%%%
%%%%%%%%%%%%%%%% FIG 1 %%%%%%%%%%%%%%%%%%

\section{Summary and Remarks}

\begin{itemize}

\item
We have developed a new perturbative framework which allows us to derive the formulas for the oscillation probabilities in matter to order $\epsilon \equiv \frac{\Delta m^2_{21}}{\dMsq} \simeq \frac{\Delta m^2_{21}}{\Delta m^2_{31}}$ in the form akin to the ones in vacuum. The correct way of decomposing the Hamiltonian into the unperturbed and perturbed parts is the key to make this property hold.  

\item
As a remarkable outcome of our machinery we have obtained the two new features of the three-flavor oscillation probabilities in matter: (i) the flavor-universal expressions of the oscillation probabilities $P(\nu_\alpha \rightarrow \nu_\beta)$ in (\ref{eq:P-albe-sec3}), and (ii) the extremely compact disappearance probability $P(\nu_e \rightarrow \nu_e)$ in (\ref{eq:P-ee-matter}). 

\item
The obvious next goal of this investigation is to extend our results to order $\epsilon^2$. Since the vacuum-like form of the oscillation probabilities hold at order $\epsilon$ and in all orders we have speculated that this property prevails to higher orders. 

\item
We have discussed in \cite{Minakata:2015gra} the issue of incorrect feature of the level crossing of the eigenvalues at the solar resonance, which appears to be a universal fault in all perturbative framework which involve $\epsilon$. I hope that we can resolve this issue in our investigation of the renormalized helio-perturbation theory to order $\epsilon^2$.

\end{itemize}

\begin{acknowledgments}
This talk is based on the collaborating work with Stephen Parke to whom the author thanks for enjoyable collaboration. He is grateful to Theory Group of Fermilab for supports and warm hospitalities during his visits. 
The author thanks Universidade de S\~ao Paulo for the great opportunity of stay under ``Programa de Bolsas para Professors Visitantes Internacionais na USP''. He is supported by Funda\c{c}\~ao de Amparo \`a Pesquisa do Estado de S\~ao Paulo (FAPESP) under grant 2015/05208-4. He thanks the support of FAPESP funding grant 2015/12505-5 which allowed him to participate in NuFact 2015.

\end{acknowledgments}

\end{document}